# Two-Particle Tight-Binding Description of Higher-Harmonic Generation in Semiconductor Nanostructures

Ulf Peschel, Thomas Lettau, Stefanie Gräfe

Friedrich Schiller University Jena

Max-Wien-Platz 1, 07743 Jena, Germany

Kurt Busch

Humboldt-Universität zu Berlin, AG Theoretische Optik & Photonik, Newtonstr. 15, 12489 Berlin, Germany, and

Max-Born-Institut, Max-Born-Str. 2A, 12489 Berlin, Germany

e-mail address: ulf.peschel@uni-jena.de

We develop a quantum mechanical theory to describe the optical response of semiconductor nanostructures with a particular emphasis on higher-order harmonic Generation. Based on a tight-binding approach we take all two-particle correlations into account thus describing the creation, evolution and annihilation of electron and holes. In the limiting case of bulk materials, we obtain the same precision as that achieved by solving the well-established semiconductor Bloch Equations. For semiconducting structures of finite extent, we also incorporate the surrounding space thus enabling a description of electron emission. In addition, we incorporate different relaxation mechanisms such as dephasing and damping of intraband currents. Moreover, the advantage of our description is that, starting from extremely precise material data as e.g., from tight-binding parameters obtained from density-functional-theory calculations, we obtain a numerical description being by far less computationally challenging and resource-demanding as comparable ab-initio approaches, e.g., those based on time-dependent density functional theory.

## 1 Introduction

Since the first observation of high-harmonic generation (HHG) in solids [1] it has become clear that its underlying physics is heavily determined by the band structure of the crystal lattice [2, 3] resulting in interesting new phenomena such as dynamical Bloch oscillations [4, 5]. This complex electron-hole dynamics in dielectrics leaves its traces in the generated spectra [6, 7]. Several theoretical approaches have been applied to describe this phenomenon, such as those based on the time-dependent Schrödinger equation (TDSE) [8] or time-dependent density-functional theory (TDDFT) [9]. In contrast to these computationally extremely challenging approaches, the semiconductor Bloch equations (SBEs) take into account the full electron-hole evolution on the level of two-particle correlations [10, 11] and have proven to be extremely successful with respect to the modeling of the bulk response. However, as SBEs are based on a (periodic) Bloch-wave representation in reciprocal space, on first sight they appear to be unsuitable for a description of quantum systems with finite spatial extent.

Several attempts have been made to efficiently simulate HHG in non-periodic semiconductor structures as it would e.g., be relevant for quantum dots. McDonald et al. have modeled HHG in semiconducting nanowires using a single-particle approach based on a frozen valence-band approximation [12]. They claimed that confinement may result in an effective enhancement mainly due to a change of the density of states. Also, HHG on localized impurities has been studied, again on the basis of a single-particle picture [13]. Explicit spatial dependencies have also to be taken into account to describe the effect of random lattice distortions [14, 15]. The authors of those works utilized a Wannier-function based tight-binding description, which not only included single-particle transport, but also interband transitions. They could demonstrate that the presence of random lattice distortions leads to HHG spectra with clean harmonic peaks, which otherwise can only be obtained for extremely short dephasing times in calculations based on SBEs.

In this work, we follow and extend this concept and develop a quantum mechanical time-dependent tight-binding model for electrons and holes enabling us to describe of the optical response of semiconductor nano structures such as e.g., quantum dots or wires. To this end we transfer the SBEs from reciprocal to real space while keeping the same precision, at least for infinite structures. Such perturbative approach results in two-particle wave functions describing the space-resolved generation and propagation of electrons and holes. It also allows to incorporate arbitrary semiconductor band structures that are obtained via density-functional theory (DFT) calculations. This tight-binding approach allows us to model interfaces as well as electron emission to the surrounding space. The derived equations lead to computationally efficient codes, allowing to simulate the nonlinear optical response of semiconducting nanostructures on a conventional laptop computer.

Here, we illustrate the basic features of the general theory for the simplest case. We restrict ourselves to a one-dimensional two-band description, taking into account only a single valence and a single conduction band. Although the actual 3D structure of e.g., a real semiconductor wire or even a dot is not adequately reflected by this treatment, it is straightforward to extend our approach to three dimensions and any semiconductor having (usually) more relevant valence bands.

The paper is structured as follows: We first define the basis of our tight-binding approach by discussing the single-particle system in the bulk. Then, we perform a second quantization towards a many-particle description. Starting from electron and hole creation and annihilation operators in reciprocal space, we derive the tight-binding Hamiltonian of the semiconductor in second quantization including light-matter interaction. As electrons may leave the nanostructure, we also include the space surrounding the dot into our approach. With the complete Hamiltonian in our hands, we then proceed to define the optical polarization and formulate a set of coupled differential equations for its evolution. Finally, we apply the newly developed scheme to our model system that resembles a CdSe quantum wire. A summary concludes the paper, and a collection of material properties and numerical parameters can be found in the appendix.

## 2 The Single-Particle System in the Bulk

*a) Eigenfunctions and Eigen Energies*

All formulas are expressed in atomic units, i.e., we set $e = 1$, $\hbar = 1$, $m_e = 1$, $\varepsilon_0 = 1$ and we restrict ourselves to a one-dimensional representation (for a straightforward generalization towards three dimensions see e.g., [16] and references therein).

We start from the single particle eigenfunctions of the unperturbed bulk crystal. In the real-space representation, the corresponding Bloch waves $|\psi_k^\eta\rangle$ of a certain band $\eta$ and Bloch vector $k$ $-\frac{\pi}{a} \leq k \leq \frac{\pi}{a}$ read as

$$\langle x | \psi_k^\eta \rangle = \psi_k^\eta(x) = u_k^\eta(x) \exp(ikx) \; , \tag{1}$$

where $u_k^\eta(x)$ is the lattice periodic part $u_k^\eta(x+a) = u_k^\eta(x)$ and $a$ denotes the size of the elementary cell. These Bloch-waves $|\psi_k^\eta\rangle$ fulfill the orthogonality relation

$$\langle \psi_{k'}^{\eta'} | \psi_k^\eta \rangle = \int_{-\infty}^{\infty} dx \, \psi_{k'}^{\eta'}(\vec{r})^* \psi_k^\eta(\vec{r}) = \delta_{\eta\eta'} \delta(k-k'). \tag{2}$$

Their phase can be chosen such that the symmetry relation $u_{-k}^\eta(x) = u_k^\eta(x)^*$ is fulfilled. The energies $\varepsilon_\eta(k)$ are eigenvalues of the single-particle Hamiltonian $\hat{\mathbf{H}}_0$, i.e.,

$$\hat{\mathbf{H}}_0 | \psi_k^\eta \rangle = \varepsilon_\eta(k) | \psi_k^\eta \rangle \; , \tag{3}$$

and have the Fourier expansion

$$\varepsilon_\eta(k) = \sum_{m=-\infty}^{\infty} \varepsilon_m^\eta \exp(imak). \tag{4}$$

As the band structure is symmetric and real, $\varepsilon_m^\eta = \varepsilon_m^{\eta*} = \varepsilon_{-m}^\eta$ holds. For the considered case of a dielectric with direct transition at $k=0$, there is a gap in the energy spectrum defined as $\varepsilon_{\text{gap}} = \sum_{m=-\infty}^{\infty} (\varepsilon_m^c - \varepsilon_m^v)$. In what follows, we consider a two-band model and define the energy zero in the middle of the gap between the valence and the conduction band.

The Fourier coefficients of the bands are related to the experimentally well-accessible effective masses given by $\frac{1}{m_\eta} = \frac{\partial^2}{\partial k^2}\varepsilon_\eta(k)\Big|_{k=0} = -2\sum_{n=1}^{\infty}(na)^2 \varepsilon_n^\eta$. Restricting ourselves to the zeroth and the first Fourier component, we obtain within the two-band model a cosine-shaped band structure as $\varepsilon_\eta(k) \approx \frac{\varepsilon_{gap}}{2} - \frac{1}{m_\eta a^2}\cos(ak)$, which will allow us later to a restrict to nearest neighbor interactions.

*b) Transition Dipole Elements between Conduction and Valence Band*

The transition dipole matrix elements between conduction and valence band are

$$D(k) = \int_{-a/2}^{a/2} dx\, u_k^{c*}(x)\, x\, u_k^v(x), \tag{5}$$

with the Fourier expansion

$$D(k) = \sum_{m=-\infty}^{\infty} D_m \exp(imak), \tag{6}$$

and $D_{-m} = D_m^*$ caused by time-inversion symmetry of the Bloch waves (see the phase choice with respect of an inversion of $k$ discussed below Eq. (2)).

Note that above definition (5) is rarely used because of the ambiguity of the definition of the elementary cell in a periodic lattice. In DFT calculations one usually determines respective momentum elements first which are later used to derive the required dipole elements (for a detailed discussion see [17, 18]). But this has no effect on the Fourier expansion (6) and the resulting symmetries.

We again restrict to the simplest case $m = 0$, i.e., a completely local optical transition without any immediate transverse transport.

*c) Tight-Binding Approach*

A maximally localized or Wannier state $|\varphi_n^\eta\rangle$ of band $\eta$ on site $n$ is constructed by a superposition of Bloch waves and reads in spatial representation as

$$\langle x|\varphi_n^\eta\rangle = \varphi_n^\eta(x) = \sqrt{\frac{a}{2\pi}}\int_{-\pi/a}^{\pi/a} dk\, \psi_k^\eta(x+na) = \sqrt{\frac{a}{2\pi}}\int_{-\pi/a}^{\pi/a} dk\, u_k^\eta(x)\exp[ik(x+na)] \tag{7}$$

where the phase of the Bloch waves is chosen such that maximum localization of the tight-binding states $\varphi_n^\eta(x)$ is obtained. As Bloch waves and Wannier states are connected by a unitary transformation, the Wannier functions obey the orthogonality relation

$$\int_{-\infty}^{\infty} dx\, \varphi_{n'}^{\eta'}(x)^* \varphi_n^\nu(x) = \delta_{\eta'\eta}\delta_{nn'}, \tag{8}$$

where $\delta_{pq}$ is the Kronecker symbol.

Different tight-binding states are solely shifted with respect to each other by multiples of the lattice period according to

$$\varphi_{n+m}^{\eta}(x) = \int_{-\pi/a}^{\pi/a} dk \, u_k^{\eta}(x) \exp\left[ik(x+na+ma)\right] = \varphi_n^{\eta}(x+ma). \tag{9}$$

## 3 Many-Particle Approach

*a) Electron Creation and Annihilation Operators of Bloch States*

In the equilibrium state $|0\rangle$ of the system, thermal excitations of conduction-band electrons can be neglected. Hence, initially all single-particle states of the valence band are occupied by electrons and all conduction-band states are empty. The creation of an electron with a certain quasi-momentum $k$ in state $|\psi_k^c\rangle$ of the conduction-band is described by the creation operator $\hat{\Psi}_k^{c+}$. It is accompanied by the removal of an electron from the valence band, a process which is described by the application of a fermionic annihilation operator $\hat{\Psi}_k^v$. Here, we have already assumed that momentum conservation holds which is typical for optical excitations. The resulting creation of an excited state $|*\rangle_k$ is thus expressed as $|*\rangle_k = \hat{\Psi}_k^{c+}\hat{\Psi}_k^v|0\rangle$.

Those creation and annihilation operators are adjoint to each other $\hat{\Psi}_k^{\eta+} = \left(\hat{\Psi}_k^{\eta}\right)^+$ and obey the anti-commutator $\left[\hat{A}, \hat{B}\right]_+ = \hat{A}\hat{B} + \hat{B}\hat{A}$ relations $\left[\hat{\Psi}_k^{\eta}, \hat{\Psi}_{k'}^{\eta'}\right]_+ = \left[\hat{\Psi}_k^{\eta+}, \hat{\Psi}_{k'}^{\eta'+}\right]_+ = 0$ and $\left[\hat{\Psi}_k^{\eta}, \hat{\Psi}_{k'}^{\eta'+}\right]_+ = \hat{\Psi}_k^{\eta}\hat{\Psi}_{k'}^{\eta'+} + \hat{\Psi}_{k'}^{\eta'+}\hat{\Psi}_k^{\eta} = \delta_{\eta\eta'}\delta(k-k')$, where the number operator $\hat{\Psi}_k^{\eta+}\hat{\Psi}_k^{\eta}$ appears.

*b) Introduction of Holes*

It is convenient to rather count holes instead of electrons in the valence-band. Hence, new electron and hole creation and annihilation operators are introduced as $\hat{\Psi}_k^v = \hat{h}_{-k}^+$ and $\hat{\Psi}_k^{c+} = \hat{e}_k^+$ and the momentum conserving creation of an electron-hole pair in now reads as, $|*\rangle_k = \hat{e}_k^+ \hat{h}_{-k}^+|0\rangle$.

While anticommutator relations and conduction band energies remain unchanged, hole energies feature a sign change as compared with valence band energies, i.e., while $\varepsilon_e(k) = \varepsilon_c(k)$ it is $\varepsilon_h(k) = -\varepsilon_v(k)$.

*c) Many-Particle Hamiltonian of the Bulk Solid in the Bloch Basis*

The simplest form of the bulk many-particle Hamiltonian in the Bloch-wave basis within our approximations (two band model, one-dimensional description, no Coulomb interaction) reads as

$$\hat{H}^b = \hat{H}_0^b + \hat{H}_{\text{inter}}^b + \hat{H}_{\text{intra}}^b. \tag{10}$$

$\hat{\mathbf{H}}_0^b$ accounts for the energies of the bulk semiconductor as defined in (3) and, in terms of creation and annihilation operators, read as

$$\hat{\mathbf{H}}_0^b = \int_{-\pi/a}^{\pi/a} dk \left[ \varepsilon_e(k) \hat{\mathbf{e}}_k^+ \hat{\mathbf{e}}_k + \varepsilon_h(k) \hat{\mathbf{h}}_k^+ \hat{\mathbf{h}}_k \right]. \tag{11}$$

$\hat{\mathbf{H}}_{\text{inter}}^b$ describes the optical transitions between bands under the action of a time-dependent electrical field $E(t)$ as

$$\hat{\mathbf{H}}_{\text{inter}}^b = E(t) \int_{-\pi/a}^{\pi/a} dk \left[ D(k) \hat{\mathbf{e}}_k^+ \hat{\mathbf{h}}_{-k}^+ + D(k)^* \hat{\mathbf{h}}_{-k} \hat{\mathbf{e}}_k \right] \tag{12}$$

The optical field also accelerates charge carriers within their respective bands, an effect, which is represented by

$$\hat{\mathbf{H}}_{\text{intra}}^b = iE(t) \int_{-\pi/a}^{\pi/a} dk \left( \hat{\mathbf{e}}_k^+ \frac{\partial}{\partial k} \hat{\mathbf{e}}_k - \hat{\mathbf{h}}_k^+ \frac{\partial}{\partial k} \hat{\mathbf{h}}_k \right). \tag{13}$$

The Hamiltonian introduced in (10) leads to the well-known SBEs [10, 11]. In what follows, we will refrain from using this Hamiltonian directly, but rather will apply a tight-binding approach which will lead to evolution equations that are as accurate as the SBEs when it comes to the description of the bulk.

*d) Tight-Binding Approach*

Tight-binding creation and annihilation operators of electrons and holes are derived from the respective operators in the Bloch basis using the unitary transformation (7), similar to what has been done for the single-particle wave functions:

$$\hat{\mathbf{e}}_n = \sqrt{\frac{a}{2\pi}} \int_{-\pi/a}^{\pi/a} dk \, \hat{\mathbf{e}}_k \exp(ikna) \quad \text{and} \quad \hat{\mathbf{h}}_n = \sqrt{\frac{a}{2\pi}} \int_{-\pi/a}^{\pi/a} dk \, \hat{\mathbf{h}}_k (ikna) \exp(ikna). \tag{14}$$

The associated inverse transformation

$$\hat{\mathbf{e}}_k = \sqrt{\frac{a}{2\pi}} \sum_n \hat{\mathbf{e}}_n \exp[-ikna] \quad \text{and} \quad \hat{\mathbf{h}}_k = \sqrt{\frac{a}{2\pi}} \sum_n \hat{\mathbf{h}}_n \exp[-ikna], \tag{15}$$

is based on the relation $\sum_n \exp(inka) = \frac{2\pi}{a} \delta(k)$ for $-\frac{\pi}{a} < k \leq \frac{\pi}{a}$.

Again, creation and annihilation operators are adjoint counterparts, i.e., $\hat{\mathbf{e}}_n^+ = (\hat{\mathbf{e}}_n)^+$ and $\hat{\mathbf{h}}_n^+ = (\hat{\mathbf{h}}_n)^+$. Hence, an excited state $|*\rangle_n$ consisting of e.g., an electron and a hole at site $n$ is created from the equilibrium state by applying the corresponding creation operators $|*\rangle_n = \hat{\mathbf{e}}_n^+ \hat{\mathbf{h}}_n^+ |0\rangle$.

As the above transformation is unitary, the respective anti-commutator and orthogonality relations result in

$$\left[\hat{\mathbf{e}}_n, \hat{\mathbf{e}}_{n'}\right]_+ = \left[\hat{\mathbf{h}}_n, \hat{\mathbf{h}}_{n'}\right]_+ = \left[\hat{\mathbf{e}}_n, \hat{\mathbf{h}}_{n'}\right]_+ = \left[\hat{\mathbf{e}}_n, \hat{\mathbf{h}}_{n'}^+\right]_+ = 0 \quad \text{and} \quad \left[\hat{\mathbf{e}}_n, \hat{\mathbf{e}}_{n'}^+\right]_+ = \delta_{nn'}, \quad \left[\hat{\mathbf{h}}_n, \hat{\mathbf{h}}_{n'}^+\right]_+ = \delta_{nn'}, \tag{16}$$

ensuring fermionic properties. The operators $\hat{N}_n^e = \hat{e}_n^+ \hat{e}_n$ and $\hat{N}_n^h = \hat{h}_n^+ \hat{h}_n$ now count electrons and holes on site *n*, respectively. Their expectation values evaluate to 0 and 1 if applied to the ground $|0\rangle$ and the set of excited states $|*\rangle_n$, respectively.

*e) Many-Particle Hamiltonian of the Bulk Solid in the Tight-Binding Basis*

Applying the inverse unitary transformation (15) to $\hat{H}_0$ defined in Eq. (11) and using the Fourier expansion of the single particle energies (4) results in

$$\hat{H}_0^{tb} = \sum_{n=-\infty}^{\infty} \sum_{m=-\infty}^{\infty} \left( \varepsilon_m^e \hat{e}_n^+ \hat{e}_{n-m} + \varepsilon_m^h \hat{h}_n^+ \hat{h}_{n-m} \right). \tag{17}$$

Note, that all higher Fourier terms $|m| \geq 1$ account for particle hopping between the sides.

In what follows we restrict to nearest neighbor interaction and use the tight-binding Hamiltonian

$$\hat{H}_0^{tb} = \sum_{n=-\infty}^{\infty} \left[ \frac{\varepsilon_{\text{gap}}}{2} \left( \hat{e}_n^+ \hat{e}_n + \hat{h}_n^+ \hat{h}_n \right) - \frac{1}{2m_e a^2} \left( \hat{e}_{n+1}^+ \hat{e}_n + \hat{e}_n^+ \hat{e}_{n+1} - \hat{e}_n^+ \hat{e}_n - \hat{e}_{n+1}^+ \hat{e}_{n+1} \right) \right.$$
$$\left. - \frac{1}{2m_h a^2} \left( \hat{h}_{n+1}^+ \hat{h}_n + \hat{h}_n^+ \hat{h}_{n+1} - \hat{h}_n^+ \hat{h}_n + \hat{h}_{n+1}^+ \hat{h}_{n+1} \right) \right] \tag{18}$$

with $\varepsilon_{\text{gap}} = \varepsilon_0^e + \varepsilon_1^e + \varepsilon_0^h + \varepsilon_1^h$.

We now apply the same procedure to the interband transitions, Eq. (12), and employ the Fourier decomposition of the dipole matrix elements, Eq. (6), resulting in

$$\hat{H}_{\text{inter}}^{tb} = E(t) \sum_{n=-\infty}^{\infty} \left\{ D_0 \hat{e}_n^+ \hat{h}_n^+ + \sum_{m=1}^{\infty} \left[ \text{Re}(D_m) \hat{e}_n^+ \left( \hat{h}_{n+m}^+ + \hat{h}_{n-m}^+ \right) + i \, \text{Im}(D_m) \hat{e}_n^+ \left( \hat{h}_{n+m}^+ - \hat{h}_{n-m}^+ \right) \right] \right\} + h.c. \tag{19}$$

where *h.c.* denotes the Hermitian conjugate. Hence, upon an optical excitation, a *k*-dependent dipole matrix element causes transverse transport. It can be symmetric in case of a real dipole element or may have an antisymmetric component for a non-vanishing imaginary part.

As a further assumption, we restrict ourselves to spatially local transitions using an interband Hamiltonian as

$$\hat{H}_{\text{inter}}^{tb} = E(t) D_0 \sum_{n=-\infty}^{\infty} \left( \hat{e}_n^+ \hat{h}_n^+ + \hat{h}_n \hat{e}_n \right) \tag{20}$$

Applying the inverse unitary transformation to $\hat{H}_{\text{intra}}^{tb}$ yields within the dipole approximation

$$\hat{H}_{\text{intra}}^{tb} = -E(t) \sum_{n=-\infty}^{\infty} a n \left( \hat{h}_n^+ \hat{h}_n - \hat{e}_n^+ \hat{e}_n \right) \tag{21}$$

# 4 Many-Particle Hamiltonian for a Semiconductor Nanostructure

Now, we discuss the interaction of a strong optical field with a confined system, a quantum wire or dot centered at position $x_0$. It is assumed to consist of semiconducting material in a range $|x - x_0| \leq L/2$, which is embedded in free space, either vacuum or a material without any crystalline structure, e.g., a liquid.

*a) Tight-Binding Description of Electron Motion*

As electrons might be able to leave the nanostructure due to the acceleration by the optical field, we have to describe the surrounding space in a manner, which is consistent with the tight-binding description of the semiconductor. Although the free space outside is continuous any numerical procedure requires its discretization it to successfully describe the motion of a quantum particle within it. Hence, a tight-binding description is justified as long as kinetic energies are described correctly, at least for relevant values of momenta.

We again start with the single-particle Hamiltonian of the electron in free space. To account for the ionization energy each electron emitted into free space is labeled with a rest energy $\varepsilon_{\text{free space}}$ as measured from the middle of the gap of the semiconductor. Within a numerical description, free space will be discretized in a regular lattice with a spacing of $dx$

$$\hat{\mathbf{H}}_0^f = \int\limits_{\text{free space}} dx \left[ \varepsilon_{\text{free space}} \psi(x)^* \psi(x) + \psi(x)^* \left( -\frac{1}{2} \frac{\partial^2}{\partial x^2} \right) \psi(x) \right]$$

$$\approx \sum_n^{\text{free space}} \left[ \varepsilon_{\text{free space}} \sqrt{dx}\, \psi(n\, dx)^* \sqrt{dx}\, \psi(n\, dx) \right.$$

$$\left. - \sqrt{dx}\, \psi(n\, dx)^* \frac{1}{2\, dx^2} \left\{ \sqrt{dx}\, \psi\left[(n+1)\, dx\right] + \sqrt{dx}\, \psi\left[(n-1)\, dx\right] - 2\sqrt{dx}\, \psi(n\, dx) \right\} \right]$$

Accordingly, in the above Hamiltonian we have replaced the continuous single-particle wave function $\psi(x)$ by a discrete set of complex amplitudes $\sqrt{dx}\, \psi(n\, dx)$ and $\sqrt{dx}\, \psi(n\, dx)^*$. Following the scheme of second quantization these amplitudes are replaced by electron annihilation $\hat{\mathbf{e}}_n$ and creation operators $\hat{\mathbf{e}}_n^+$ where the normalization is chosen such that $\hat{\mathbf{e}}_n^+ \hat{\mathbf{e}}_n$ counts the number of electrons in the space interval $\left(n - \frac{1}{2}\right) dx < x \leq \left(n + \frac{1}{2}\right) dx$.

This results in a free-space tight-binding Hamiltonian with discretized spatial derivative as

$$\hat{\mathbf{H}}_0^f = \sum_n^{\text{free space}} \left[ \varepsilon_{\text{free space}} \hat{\mathbf{e}}_n^+ \hat{\mathbf{e}}_n - \frac{1}{2\, dx^2} \left( \hat{\mathbf{e}}_n^+ \hat{\mathbf{e}}_{n+1} + \hat{\mathbf{e}}_{n+1}^+ \hat{\mathbf{e}}_n - \hat{\mathbf{e}}_n^+ \hat{\mathbf{e}}_n - \hat{\mathbf{e}}_{n+1}^+ \hat{\mathbf{e}}_{n+1} \right) \right] \tag{22}$$

Hence, within a tight-binding approximation electrons in free space and in the conduction band of a semiconductor are described in a very similar fashion (compare Eqs. (18) and (22)).

As for every numerical description also discretization of free space has to be performed carefully. Accumulated phase differences between neighboring sites must remain much smaller than π. Hence, the higher the momenta aquired by electrons outside the nanostructure, the finer the discretization (here $dx$) has to be.

In what follows we will combine bulk (18) and free space (22) versions into a global Hamiltonian

$$\hat{\mathbf{H}}_0 = \sum_{n=-\infty}^{\infty} \Big[ \varepsilon_n^e \hat{\mathbf{e}}_n^+ \hat{\mathbf{e}}_n + \varepsilon_n^h \hat{\mathbf{h}}_n^+ \hat{\mathbf{h}}_n - c_n^e \Big( \hat{\mathbf{e}}_{n+1}^+ \hat{\mathbf{e}}_n + \hat{\mathbf{e}}_n^+ \hat{\mathbf{e}}_{n+1} - \hat{\mathbf{e}}_n^+ \hat{\mathbf{e}}_n - \hat{\mathbf{e}}_{n+1}^+ \hat{\mathbf{e}}_{n+1} \Big) \\ - c_n^h \Big( \hat{\mathbf{h}}_{n+1}^+ \hat{\mathbf{h}}_n + \hat{\mathbf{h}}_n^+ \hat{\mathbf{h}}_{n+1} - \hat{\mathbf{h}}_n^+ \hat{\mathbf{h}}_n + \hat{\mathbf{h}}_{n+1}^+ \hat{\mathbf{h}}_{n+1} \Big) \Big] \tag{23}$$

with a space dependent electron energy

$$\varepsilon_n^e = \begin{cases} \dfrac{\varepsilon_g}{2} & \text{inside the nanostructure for } |na - x_0| \leq L/2 \\ \varepsilon_{\text{free space}} & \text{in free space} \end{cases}.$$

The coupling constants of electrons $c_n^e$ between sites $n$ and $n+1$ are also spatially dependent, according to

$$c_n^e = \begin{cases} \dfrac{1}{2a^2 m_e} & \text{inside the nanostructure for } |na - x_0| \leq L/2 \\ \dfrac{1}{2\,dx^2} & \text{in free space} \end{cases}.$$

In contrast, holes cannot enter free space resulting in a vanishing coupling outside the nanostructure, thus

$$c_n^h = \begin{cases} \dfrac{1}{2a^2 m_h} & \text{inside the nanostructure for } |na - x_0| \leq L/2 \\ 0 & \text{in free space} \end{cases}.$$

Although such treatment allows for a consistent description of semiconductor nanostructures in their respective environment, surface states might not be reproduced correctly and, therefore, for realistic applications of our framework, further modification of tight-binding parameters close to the interface might be required.

*b) Optical Field Action in the Nanostructure and Free Space*

Optical transitions can only happen inside the semiconductor. Therefore, $\hat{\mathbf{H}}_{\text{inter}}^{tb}$ in Eq. (20) does not need any modification except for setting the transition dipole matrix element to zero outside, resulting in

$$\hat{\mathbf{H}}_{\text{inter}} = E(t) \sum_{n=-\infty}^{\infty} d_n \Big( \hat{\mathbf{e}}_n^+ \hat{\mathbf{h}}_n^+ + \hat{\mathbf{h}}_n \hat{\mathbf{e}}_n \Big) \tag{24}$$

with

$$d_n = \begin{cases} D_0 & \text{inside the nanostructure for } |na - x_0| \leq L/2 \\ 0 & \text{in free space} \end{cases}$$

Electrons in free space are accelerated by the optical field. Hence, $\hat{\mathbf{H}}_{intra}^{tb}$ of Eq. (21) needs to be modified in order to include near-field effects as

$$\hat{\mathbf{H}}_{intra} = E(t) \sum_{n=-\infty}^{\infty} V_n \left( \hat{\mathbf{h}}_n^+ \hat{\mathbf{h}}_n - \hat{\mathbf{e}}_n^+ \hat{\mathbf{e}}_n \right). \tag{25}$$

This incorporates a spatially inhomogeneous and time-dependent effective potential $E(t)V_n$. In a quasi-static approximation, the time-dependent electric field $E(t)$ is combined with a spatially varying factor $V_n$ that accounts for local field enhancements. We assume that the overall nonlinear response is weak and that the spatial shape of the optical driving field is still determined by the linear optical response of the structure.

To demonstrate the effects of spatial dependence of the incidence field, we approximate the optical field around the ends of the nanostructure by that occurring around a spherical particle centered at $x_0$, for which analytical expressions are known for the static case [19]

$$\begin{aligned} V_n &= -\left( \frac{3\varepsilon_{\text{free space}}^R}{\varepsilon_{\text{bulk}}^R + 2\varepsilon_{\text{free space}}^R} \right)(na - x_0) & \text{inside the nanostructure} \\ V_n &= -(na - x_0) + \left( \frac{\varepsilon_{\text{bulk}}^R - \varepsilon_{\text{free space}}^R}{\varepsilon_{\text{bulk}}^R + 2\varepsilon_{\text{free space}}^R} \right) \frac{(L/2)^3}{|na - x_0|^3}(na - x_0) & \text{in free space} \end{aligned} \tag{26}$$

In the above expressions, we denote with $\varepsilon_{\text{bulk}}^R$ and $\varepsilon_{\text{free space}}^R$, respectively, the relative dielectric constants of the semiconductor material and of the surrounding space at the frequency of the incident light field.

*c) Condensed Notation of the Hamiltonian*

Taken together, the Hamiltonian $\hat{\mathbf{H}} = \hat{\mathbf{H}}_0 + \hat{\mathbf{H}}_{inter} + \hat{\mathbf{H}}_{intra}$ can be summarized as

$$\hat{\mathbf{H}} = \sum_{n=-\infty}^{\infty} \left[ \tilde{\varepsilon}_n^e \hat{\mathbf{e}}_n^+ \hat{\mathbf{e}}_n + \tilde{\varepsilon}_n^h \hat{\mathbf{h}}_n^+ \hat{\mathbf{h}}_n + d_n E(t) \left( \hat{\mathbf{e}}_n^+ \hat{\mathbf{h}}_n^+ + \hat{\mathbf{h}}_n \hat{\mathbf{e}}_n \right) - c_n^e \left( \hat{\mathbf{e}}_{n+1}^+ \hat{\mathbf{e}}_n + \hat{\mathbf{e}}_n^+ \hat{\mathbf{e}}_{n+1} \right) - c_n^h \left( \hat{\mathbf{h}}_{n+1}^+ \hat{\mathbf{h}}_n + \hat{\mathbf{h}}_n^+ \hat{\mathbf{h}}_{n+1} \right) \right] \tag{27}$$

where we have introduced time- and space-dependent electron $\tilde{\varepsilon}_n^e(t) = \varepsilon_n^e + c_{n-1}^e + c_n^e - E(t)V_n$ and hole energies $\tilde{\varepsilon}_n^h(t) = \varepsilon_n^h + c_{n-1}^h + c_n^h + E(t)V_n$ to simplify the notation for the subsequent derivation of the evolution equation of the polarization.

# 5 The Evolution of Polarization and Current within the Tight-Binding Approximation

*a) Definition of the Polarization and Currents interacting with the Optical Field*

In a classical Hamiltonian description, light-matter interaction is represented by the term $\sim \vec{P}\vec{E}$ formed by a product of the field and polarization. Hence, we can immediately identify the operator of the optical polarization by collecting all the terms that are connected with the electric field. It follows from (24) and (25) that the total polarization is

$$\hat{\mathbf{P}} = \sum_{n=-\infty}^{\infty} \left[ d_n \left( \hat{\mathbf{e}}_n^+ \hat{\mathbf{h}}_n^+ + \hat{\mathbf{h}}_n \hat{\mathbf{e}}_n \right) + V_n \left( \hat{\mathbf{h}}_n^+ \hat{\mathbf{h}}_n - \hat{\mathbf{e}}_n^+ \hat{\mathbf{e}}_n \right) \right]. \tag{28}$$

It contains the non-Hermitian interband polarizations $d_n \hat{\mathbf{h}}_n \hat{\mathbf{e}}_n$ on a site $n$ and the polarization due to charge imbalances $V_n \left( \hat{\mathbf{h}}_n^+ \hat{\mathbf{h}}_n - \hat{\mathbf{e}}_n^+ \hat{\mathbf{e}}_n \right)$ induced by macroscopic currents. The latter term is only relevant in nanostructures where charge imbalances can accumulate close to interfaces.

Also, intraband currents drive the optical field. In a one-dimensional k-space representation they are represented by [10, 11]

$$\hat{\mathbf{J}}(t) = \int_{-\pi/a}^{\pi/a} dk \left[ v_h(k) \hat{\mathbf{h}}_k^+ \hat{\mathbf{h}}_k - v_e(k) \hat{\mathbf{e}}_k^+ \hat{\mathbf{e}}_k \right] \tag{29}$$

with $v_{e/h}(k)$ being the group velocities of electrons and holes, which are defined by derivatives of the dispersion relation. For cosine shaped bands they are denoted by

$$v_{e/h}(k) = \frac{\partial}{\partial k} \varepsilon_{e/h}(k) = \frac{1}{m_{e/h} a} \sin(ak) \tag{30}$$

We now insert this expression into Eq. (29), apply transformation (15) and come to a space resolved expression for the operator of the current density as

$$\hat{\mathbf{J}}(t) = \frac{1}{2ia} \sum_n \left[ \frac{1}{m_h} \left( \hat{\mathbf{h}}_n^+ \hat{\mathbf{h}}_{n+1} - \hat{\mathbf{h}}_n^+ \hat{\mathbf{h}}_{n-1} \right) - \frac{1}{m_e} \left( \hat{\mathbf{e}}_n^+ \hat{\mathbf{e}}_{n+1} - \hat{\mathbf{e}}_n^+ \hat{\mathbf{e}}_{n-1} \right) \right] \tag{31}$$

*b) Evolution Equations of Operators*

In order to determine the optical polarization within the Heisenberg picture, we first use the Schrödinger equation to derive evolution equations for creation and annihilation operators based on the Hamiltonian of the dot $\hat{\mathbf{H}}$ displayed in Eq. (27) as

$$i \frac{d}{dt} \hat{\mathbf{e}}_n = \left[ \hat{\mathbf{e}}_n, \hat{\mathbf{H}} \right] = \hat{\mathbf{e}}_n \hat{\mathbf{H}} - \hat{\mathbf{H}} \hat{\mathbf{e}}_n = \tilde{\varepsilon}_n^e \hat{\mathbf{e}}_n - c_{n-1}^e \hat{\mathbf{e}}_{n-1} - c_n^e \hat{\mathbf{e}}_{n+1} + E(t) d_n \hat{\mathbf{h}}_n^+$$
$$i \frac{d}{dt} \hat{\mathbf{h}}_n = \left[ \hat{\mathbf{h}}_n, \hat{\mathbf{H}} \right] = \hat{\mathbf{h}}_n \hat{\mathbf{H}} - \hat{\mathbf{H}} \hat{\mathbf{h}}_n = \tilde{\varepsilon}_n^h \hat{\mathbf{h}}_n - c_{n-1}^h \hat{\mathbf{h}}_{n-1} - c_n^h \hat{\mathbf{h}}_{n+1} - E(t) d_n \hat{\mathbf{e}}_n^+ \tag{32}$$

All operators on site $n$ are coupled to those from neighboring sites. Therefore, all two-particle operators $\hat{\mathbf{n}}_{nm}^e = \hat{\mathbf{e}}_n^+ \hat{\mathbf{e}}_m$, $\hat{\mathbf{n}}_{nm}^h = \hat{\mathbf{h}}_n^+ \hat{\mathbf{h}}_m$ and $\hat{\mathbf{p}}_{nm} = \hat{\mathbf{h}}_n \hat{\mathbf{e}}_m$ are mutually coupled as

$$i \frac{d}{dt} \hat{\mathbf{n}}_{nm}^e = \left( \tilde{\varepsilon}_m^e - \tilde{\varepsilon}_n^e \right) \hat{\mathbf{n}}_{nm}^e - c_{m-1}^e \hat{\mathbf{n}}_{nm-1}^e - c_m^e \hat{\mathbf{n}}_{nm+1}^e + c_{n-1}^e \hat{\mathbf{n}}_{n-1m}^e + c_n^e \hat{\mathbf{n}}_{n+1m}^e + E(t) \left( d_m \hat{\mathbf{p}}_{mn}^+ - d_n \hat{\mathbf{p}}_{nm} \right)$$

$$i \frac{d}{dt} \hat{\mathbf{n}}_{nm}^h = \left( \tilde{\varepsilon}_m^h - \tilde{\varepsilon}_n^h \right) \hat{\mathbf{n}}_{nm}^h - c_{m-1}^h \hat{\mathbf{n}}_{nm-1}^h - c_m^h \hat{\mathbf{n}}_{nm+1}^h + c_{n-1}^h \hat{\mathbf{n}}_{n-1m}^h + c_n^h \hat{\mathbf{n}}_{n+1m}^h + E(t) \left( d_m \hat{\mathbf{p}}_{nm}^+ - d_n \hat{\mathbf{p}}_{mn} \right)$$

$$i \frac{d}{dt} \hat{\mathbf{p}}_{nm} = \left( \tilde{\varepsilon}_m^e + \tilde{\varepsilon}_n^h \right) \hat{\mathbf{p}}_{nm} - c_{m-1}^e \hat{\mathbf{p}}_{nm-1} - c_m^e \hat{\mathbf{p}}_{nm+1} - c_{n-1}^h \hat{\mathbf{p}}_{n-1m} - c_n^h \hat{\mathbf{p}}_{n+1m} + E(t) \left( d_m \delta_{nm} - d_n \hat{\mathbf{n}}_{nm}^e - d_m \hat{\mathbf{n}}_{mn}^h \right).$$

*c) Expectation Values*

As we use the Heisenberg picture for the time-dependence of the operators, the system remains in the initial or equilibrium state $|0\rangle$. Time-dependent expectation values of the polarization (28) and the intraband current (31) are, therefore, given by

$$P(t) = \langle 0|\hat{\mathbf{P}}(t)|0\rangle = \sum_{n=-\infty}^{\infty} \left[ d_n \left( \langle 0|\hat{\mathbf{p}}_{nn}|0\rangle^* + \langle 0|\hat{\mathbf{p}}_{nn}|0\rangle \right) + V_n \left( \langle 0|\hat{\mathbf{n}}_{nn}^h|0\rangle - \langle 0|\hat{\mathbf{n}}_{nn}^e|0\rangle \right) \right]$$
$$J(t) = \langle 0|\hat{\mathbf{J}}(t)|0\rangle = \frac{a}{i} \sum_{n=-\infty}^{\infty} \left[ c_n^h \left( \langle 0|\hat{\mathbf{n}}_{nn+1}^h|0\rangle - \langle 0|\hat{\mathbf{n}}_{nn-1}^h|0\rangle \right) - c_n^e \left( \langle 0|\hat{\mathbf{n}}_{nn+1}^e|0\rangle - \langle 0|\hat{\mathbf{n}}_{nn-1}^e|0\rangle \right) \right]$$
(33)

where we have expressed effective masses by respective coupling constants. Radiation is generated by accelerated carriers and is proportional to the squared Fourier transform of the second derivative of the polarization and of the first derivative of the current as

$$S(\omega) \sim \left| \omega^2 \mathrm{FT}[P(t)](\omega) + i\omega \mathrm{FT}[J(t)](\omega) \right|^2. \tag{34}$$

To determine $P$ and $J$ we must follow the evolution of expectation values of two-particle correlations, i.e.,

$$n_{nm}^e(t) = \langle 0|\hat{\mathbf{n}}_{nm}^e|0\rangle, \tag{35}$$

$$n_{nm}^h(t) = \langle 0|\hat{\mathbf{n}}_{nm}^h|0\rangle, \text{ and} \tag{36}$$

$$p_{nm}(t) = \langle 0|\hat{\mathbf{p}}_{nm}|0\rangle. \tag{37}$$

Finally, the complete set of evolution equations required to determine the HHG spectrum (34) reads as

$$i\frac{d}{dt}n_{nm}^e = \left(\tilde{\varepsilon}_m^e - \tilde{\varepsilon}_n^e\right)n_{nm}^e - c_{m-1}^e n_{nm-1}^e - c_m^e n_{nm+1}^e + c_{n-1}^e n_{n-1m}^e + c_n^e n_{n+1m}^e + E(t)\left(d_m p_{mn}^* - d_n p_{nm}\right), \tag{38}$$

$$i\frac{d}{dt}n_{nm}^h = \left(\tilde{\varepsilon}_m^h - \tilde{\varepsilon}_n^h\right)n_{nm}^h - c_{m-1}^h n_{nm-1}^h - c_m^h n_{nm+1}^h + c_{n-1}^h n_{n-1m}^h + c_n^h n_{n+1m}^h + E(t)\left(d_m p_{nm}^* - d_n p_{mn}\right), \tag{39}$$

$$i\frac{d}{dt}p_{nm} = \left(\tilde{\varepsilon}_m^e + \tilde{\varepsilon}_n^h\right)p_{nm} - c_{m-1}^e p_{nm-1} - c_m^e p_{nm+1} - c_{n-1}^h p_{n-1m} - c_n^h p_{n+1m} + E(t)\left(d_m \delta_{nm} - d_n n_{nm}^e - d_m n_{mn}^h\right). \tag{40}$$

## 6 Damping

The properties of real systems are critically determined by fast damping processes which go far beyond a two-particle description. In SBEs an artificial phase relaxation time $T_2$ is introduced and ensures optical spectra with well-separated harmonic peaks with Lorentzian line shape [10, 11]. In order to introduce those relaxation times, we modify the energies within the nanostructure according to

$$\tilde{\varepsilon}_n^e(t) = \varepsilon_n^e - \frac{i}{2T_2} + c_{n-1}^e + c_n^e - E(t)V_n \text{ and } \tilde{\varepsilon}_n^h(t) = \varepsilon_n^h - \frac{i}{2T_2} + c_{n-1}^h + c_n^h + E(t)V_n.$$

Note, that such a change influences the evolution equation (40) of the polarization only, but not those of the carrier concentrations.

Also, intraband or Ohmic currents are damped with a characteristic time $T_j$. In momentum space such dissipative processes are described by a forced relaxation of carriers towards the bottom of respective bands.

According to Eq. (33) imaginary parts of off-diagonal elements $n_{nn-1}^{e/h}$ and $n_{nn+1}^{e/h}$ represent Ohmic currents in a space-resolved representation. If those shall decay, a damping of these components via additional terms of the form $-\frac{1}{T_j}\left(n_{nm}^{e/h} - n_{mn}^{e/h}\right)$ in Eqs. (38) and (39) and $-\frac{1}{T_j}\left(p_{nm} - p_{mn}\right)$ in Eq. (40) must be realized inside the semiconductor. These damping terms become active only for pronounced asymmetries of respective two-particle correlation functions.

We assume damping to happen inside the nanostructure only, resulting in space dependent relaxation times as

$$T_n^{p/j} = \begin{cases} T_{2/j} & \text{inside the nanostructure for } |na - x_0| \leq L/2 \\ \infty & \text{in free space} \end{cases}$$

Although such description of relaxation phenomena is the simplest possible approach, the respective relaxation times are not well known and vary considerably from system to system. Hence, a more involved treatment would also require a deeper investigation of relaxation by its own.

## 7 Final Set of Evolution Equations

The final set of evolution equations, which have to be solved numerically reads as

$$i\frac{d}{dt}n_{nm}^e = \left[\varepsilon_m^e - \varepsilon_n^e + c_{m-1}^e + c_m^e - c_{n-1}^e - c_n^e - E(t)(V_m - V_n)\right]n_{nm}^e - \frac{i}{2}\left(\frac{1}{T_n^j} + \frac{1}{T_m^j}\right)\left(n_{nm}^e - n_{mn}^e\right) \\ - c_{m-1}^e n_{nm-1}^e - c_m^e n_{nm+1}^e + c_{n-1}^e n_{n-1m}^e + c_n^e n_{n+1m}^e + E(t)\left(d_m p_{mn}^* - d_n p_{nm}\right) \quad (41)$$

$$i\frac{d}{dt}n_{nm}^h = \left[\varepsilon_m^h - \varepsilon_n^h + c_{m-1}^h + c_m^h - c_{n-1}^h - c_n^h + E(t)(V_m - V_n)\right]n_{nm}^h - \frac{i}{2}\left(\frac{1}{T_n^j} + \frac{1}{T_m^j}\right)\left(n_{nm}^h - n_{mn}^h\right) \\ - c_{m-1}^h n_{nm-1}^h - c_m^h n_{nm+1}^h + c_{n-1}^h n_{n-1m}^h + c_n^h n_{n+1m}^h + E(t)\left(d_m p_{nm}^* - d_n p_{mn}\right) \quad (42)$$

$$i\frac{d}{dt}p_{nm} = \left[\varepsilon_m^e + \varepsilon_n^h - \frac{i}{2}\left(\frac{1}{T_n^p} + \frac{1}{T_m^p}\right) + E(t)(V_n - V_m) + c_{m-1}^e + c_m^e + c_{n-1}^h + c_n^h\right]p_{nm} \\ - \frac{i}{2}\left(\frac{1}{T_n^j} + \frac{1}{T_m^j}\right)(p_{nm} - p_{mn}) - c_{m-1}^e p_{nm-1} - c_m^e p_{nm+1} - c_{n-1}^h p_{n-1m} - c_n^h p_{n+1m} \\ + E(t)\left(d_m \delta_{nm} - d_n n_{nm}^e - d_m n_{mn}^h\right) \quad (43)$$

Here, all computed quantities, namely $n_{nm}^e$, $n_{nm}^h$ and $p_{nm}$ vanish before the arrival of the pulse.

# 8   Numerical Results: HHG in a CdSe Nanostructure

The above-derived set of equations (41)–(43) is capable of describing the optical response of a semiconductor nanostructure within the limits set by the underlying approximations. Here, study HHG in a nanowire made from CdSe, the material parameters of which are well known (see Appendix). Numerical integration of the differential equations (41)–(43) is performed using a standard adaptive Runge-Kutta scheme [20]. Modeling of a nanostructure of 2 nm length embedded in 15 nm of surrounding space and excited by a 100 fs pulse required roughly 30 minutes of calculation on a single core (no multi-threading) of a 3.5 GHz i7 processor with 8 GByte memory.

We investigate two different quantum wires, one of 2.2 nm and one of 50 nm length containing 5 and 104 sites, respectively. We demonstrate that both, quantum confinement and the linear optical response of semiconducting nanostructures is reproduced. To probe the linear spectral response, the system is excited with a weak delta-like pulse (non-oscillating Gaussian shape, centered at $t$=0, $W_{FWHM}$=0.1fs, peak amplitude 0.001 V/nm) and follow the evolution of the microscopic expectation values (41), (42), and (43). As the current $J$ due to the absence of excited carriers is negligible, the resulting polarization, Eq. (33), is then Fourier-transformed to obtain the complex linear polarizability $p$. The imaginary part of this polarizability, as displayed in Fig.1 per site, is linked to linear absorption. Not surprisingly, the 50 nm wire (blue line) behaves like a one-dimensional bulk solid, featuring a broad quasi-structureless absorption band, which is limited to the spectral range of the interband transition between the two cosine-shaped bands (blue line). We also see the typical enhancement at the band edge as it is expected for a one-dimensional structure. Due to the absence of Coulomb interaction in our model the otherwise well-pronounced exciton peak is missing. In contrast, the linear optical response of the 2.2 nm structure (red line) is determined by quantum confinement causing the spectrum to be structured with localized absorption features corresponding to the number of real-space sites involved.

After having validated the linear optical properties, we next investigate the systems' nonlinear optical response. Interaction intense optical pulses (carrier wavelength 4000 nm, $W_{FWHM}$=100 fs, peak amplitude in free space 2.5 V/nm) results in complex carrier dynamics including the excitation of electrons and holes as displayed for the tiny wire (2.2 nm length or 5 sites) in Fig. 2. The evolution is dominated by Rabi oscillations, but due to the finite relaxation times ($T_2$=$T_j$=10 fs) a net-generation occurs so that electrons and holes remain after the pulse has passed. In what follows we will investigate how this carrier dynamics affects HHG.

We first model HHG for the large (50 nm) and the tiny (2.2 nm) wire, where we expect bulk behavior for the first and confinement effects for the second system (see Fig. 3). Indeed, the spectra of the large wire (see Fig. 2, blue line) are similar to those obtained on the basis of SBEs for bulk semiconductors. They do not coincide as three-dimensional systems have usually been studied in the literature. Still, we find a characteristic plateau region, which stretches from the band gap (1.75 eV) up to the maximum possible energy of an interband transition (within our model roughly 8 eV). In contrast, the HHG spectrum of the tiny structure is obviously affected by confinement effects. It is orders of magnitude weaker, noticeably distorted and decays more rapidly than the bulk spectrum. Even, if scaled with the squared number of sites, the power in almost all harmonic orders of the large (104 sites) structure is considerably higher than that emitted by the tiny (5 sites) wire. Moreover, the harmonic orders of the tiny structure also drop-off much quicker such that already the seventh order is more than 300 times weaker than the first one. In the case of the large structure only the 17$^{th}$ order is comparably reduced compared with the first one. Only for high frequencies the spectrum of the small wire recovers and stretches even to higher power levels than that of the bulk-like structure. However, the associated energies correspond to the upper edge of our conduction band. Hence, the latter behavior is highly affected by the chosen cosine band structure of our reduced model.

Taken together, two reasons for this modification of HHG in nanostructures can be identified. On the one hand, there is a considerable change of the energy structure, which not only affects the linear optical properties (see Fig. 1). On the other hand, in tiny nanostructures electrons and holes constantly interact with the interface, an effect which is also missing in the bulk. Our model allows us two disentangle these two distributions.

First, we mimic complete confinement by increasing the free space energy $\varepsilon_{\text{free space}}$ in Eq. (22) by a factor of six. This immediately causes a quick drop-off of HHG spectra as demonstrated in Fig. 4 (lower black line). Hence, the transformation of an otherwise continuous spectrum into a few resonances impedes HHG considerably. Thus, we can assign this to be the predominant reason for the lower yield of the tiny wire. The overall reduction of HHG is even stronger than in a nanostructure with realistic walls (cf. black and red lines in Fig. 3). Consequently, the properties of the boundaries, which are constantly probed by accelerated carriers have a strong impact on HHG in nanostructures.

For a finite binding potential, the spatial shape of the wavefunction is not considerably changed compared with the eigenstates of a box potential. But the optical field shakes the carrier distribution and electrons may exit into free space when driven by the electric field (cf. Fig. 2). Emitted electrons experience the optical near field which can be considerably stronger than the electrical field inside the semiconductor (for a sphere, see Eq. (26)). To quantify this effect, we assume that the dielectric constant of the space surrounding the nanostructure coincides with that of the semiconductor. This keeps the optical field in the surrounding space constant at its bulk value within the entire simulation domain. This homogenization of the electrical field has a pronounced negative effect on HHG as demonstrated in Fig. 3 (green line). We do not only observe a strong reduction of yield, but also a considerably reduced cut-off of the spectrum. Consequently, we conclude that the near-field enhancement amplifies HHG. Thus, optically accelerated electrons can be used to probe electrical near fields thereby connecting the wire's quantum dynamics with electrodynamics.

Periodically driven particles that interact with boundaries tend to move chaotically, as is well-known from classical mechanics. In our nanostructure, this behavior is partially suppressed by damping, a feature which we have included into our simulations via the finite phase relaxation time $T_2$ and the finite intraband current damping time $T_j$ (see Eqs. (41)–(43)). Still, traces of the aforementioned irregular motion are seen e.g., in the evolution of the center of gravity of our carrier distribution. They become more apparent if we completely manually switch off the damping of the interband currents (compare upper left and right parts of Fig. 5).

The interplay of such quasi-chaotic behavior and damping has a profound impact on HHG (see lower plot of Fig. 5). If all relaxation times are set to infinity, Rabi oscillations prevail, and all carriers are periodically generated and completely reabsorbed. A few discrete higher harmonics can still be found (see blue curve in the lower plot of Fig. 5), but carrier motion and all frequency components generated by it are missing. This changes if dephasing is added so that electrons and holes are continuously generated. Due to the ongoing field-induced acceleration and the interaction with the interfaces, these carriers undergo a quasi-chaotic motion and, as a consequence, a large plateau of newly generated frequencies emerges (see black curve in lower plot of Fig. 5). Damping of interband currents again quenches this quasi-chaotic motion (compare the two upper plots of Fig. 5), but now regular spectral peaks evolve (see red line in the lower plot of Fig. 5).

## 9 Summary and Conclusions

In conclusion, we have transferred the successful approach of SBEs to describe HHG in bulk matter from reciprocal to real space using a tight-binding approach. We describe intense-field-driven semiconductor nanostructures with the same precision as the bulk material gaining detailed insight into the field-driven electron-hole dynamics and the evolution of two-particle correlations of finite systems, in particular, in the vicinity of interfaces.

Moreover, we have applied our approach to selected science cases – a one-dimensional two-band nanoscale semiconductor embedded in free space that interacts with a strong light field and generating higher-order harmonic radiation. Our approach allows us not only to reliably and computationally efficiently simulate light-driven carrier dynamics inside nanoscale semiconductors but also includes possible strong-field induced emission of electrons into the environment. We have shown that quantum confinement may reduce the HHG efficiency and that the interaction between the excited carriers and the boundaries leaves significant traces in the higher-harmonic spectra. In particular, accelerated and partially transmitted electrons constantly probe the optical near field around the nanostructure leading to emitted spectra, which depend on the strength of the field enhancement around the nanoparticle. Moreover, we also have incorporated different possible relaxation phenomena, specifically phase relaxation and damping of intraband currents. The shape and extent of the generated spectra depend sensitively on the choice of respective relaxation times.

Although we have here demonstrated our scheme for a simple one-dimensional case, it is straightforward to extend it into various directions. Three-dimensional objects consisting of semiconductors with many and potentially anisotropic bands can be incorporated using tight-binding parameters derived from state-of-the-art DFT codes. As the carrier evolution in each unit cell of the semiconductor crystal is represented by a few complex numbers only, the resulting growth of numerical complexity can still be handled. Even for three-dimensional structures it will be considerably lower than that required for standard time dependent DFT calculations. It has to be emphasized that our approach is capable of incorporating quasi-particles consisting of pairs of particles such as excitons. In the present work, we have not included these excitonic effects yet as the applied optical fields have been much stronger than the expected internal forces that would form the excitons. Nonetheless, an inclusion of Coulomb interaction into our framework is straightforward and would not increase the complexity of our code considerably.

We acknowledge funding by the DFG in the framework of SFB 1375 NOA.

# 10 Appendix

## 10.1 Parameters of a CdSe Quantum Dot used in Simulations

a) Size of the Unit Cell of CdSe

$\tilde{a} = 4.3\text{A}, \tilde{c} = 7\text{A} \Rightarrow \bar{a} = \sqrt[3]{\tilde{a}^2\tilde{c}} = 4.8\text{A}$ normalized $a = 9.057$

b) Coupling

electrons inside the dot $m_e = 0.13 \Rightarrow c_n^e = \dfrac{1}{2a^2 m_e} = 0.04689$

electrons in free space: $c_n^e = \dfrac{1}{2\,dx^2}$ with $dx$: space discretization in normalized units

holes inside the dot: $m_h \approx 0.8 \Rightarrow c_n^h = \dfrac{1}{2a^2 m_h} = 0.00762$

c) Energies

gap: $E_{gap} = 1.75\text{eV} \Rightarrow \varepsilon_{gap} = 0.063$

free space: $E_{valence\ band} = -6.69\text{eV} \Rightarrow \varepsilon_{free\ space} = \dfrac{-E_{gap}/2 - E_{valence\ band}}{E_{Hartree}} = 0.214$

d) Dipole Matrix elements

momentum matrix element: $\langle p_k \rangle = \displaystyle\int_{-a/2}^{a/2} dx\, u_k^{c*}(x) \dfrac{1}{i}\dfrac{\partial}{\partial x} u_k^v(x)$

$\dfrac{\langle p_k \rangle^2}{2 m_0} = 20 \pm 1\,\text{eV}$ (according to C. Hermann and C. Weisbuch, Physical Revie B **15**, 823 (1977))

$$\tilde{d} \approx \frac{\hbar}{m_0 E_{gap}} \langle \tilde{p} \rangle = 5 \text{A} \Rightarrow D_0 \approx 9.4$$

f) Relative Dielectric Constant

inside the dot: $\varepsilon_{\text{CdSe}}^{R} \approx 6$

outside: $\varepsilon_{\text{free space}}^{R} = 1$

## 10.2 Calculation of HHG spectra

The polarization $P(t) = \sum_{n=-\infty}^{\infty} \left[ d_n \left( p_{nn} + p_{nn}^* \right) + V_n \left( n_{nn}^h - n_{nn}^e \right) \right]$ is determined according to Eq. (33) during each time step. The computations are continued until all interband polarizations and interband currents have decayed (usually 3 times the pulse width (FWHM)). The final data file $P(t)$ was artificially extended to 20 times its original length by zero-padding in order to increase the frequency resolution. Then, the spectral intensity was determined according to Eq. (34). Finally, fast oscillations of this power spectrum were filtered out by making a convolution with a Gaussian filter function with a width of a fifth of the fundamental frequency, thus simulating a spectrometer with a limited spectral resolution.

## 12 Figure Captions

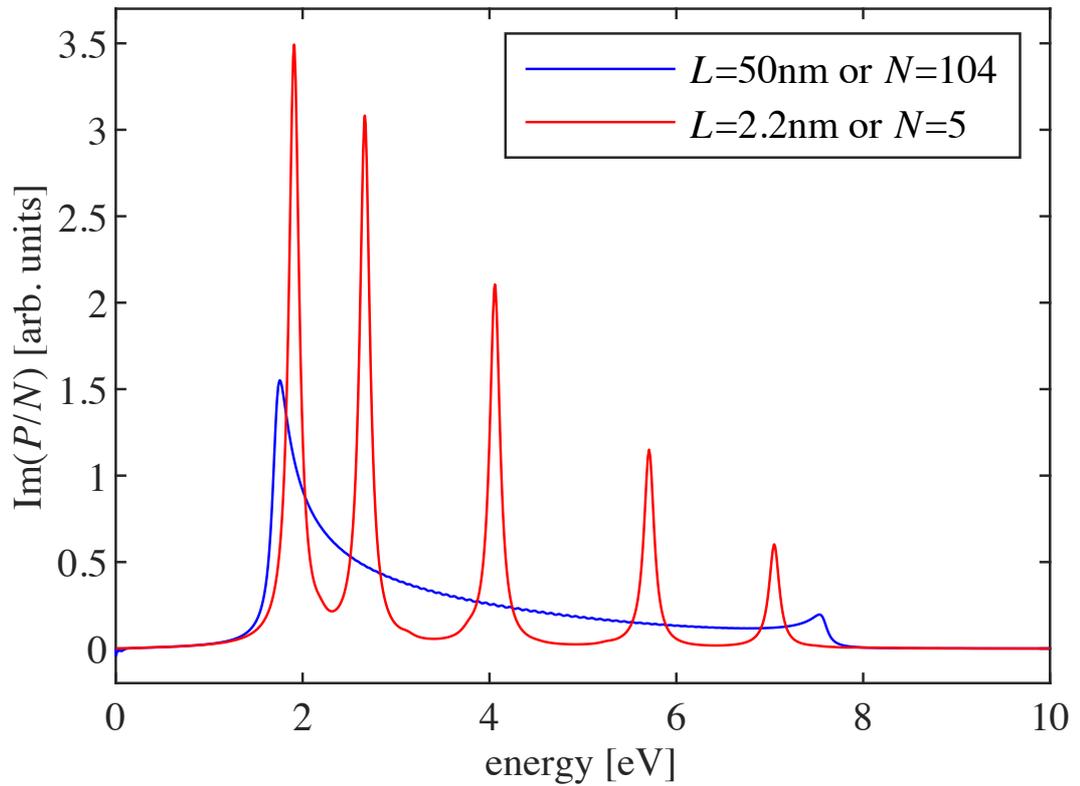

Fig. 1: Linear optical properties of nanostructures under investigation. Imaginary parts of polarizability per site are displayed for a large bulk-like nanowire and a small one, for which quantum confinement plays an essential role (for parameters see Appendix).

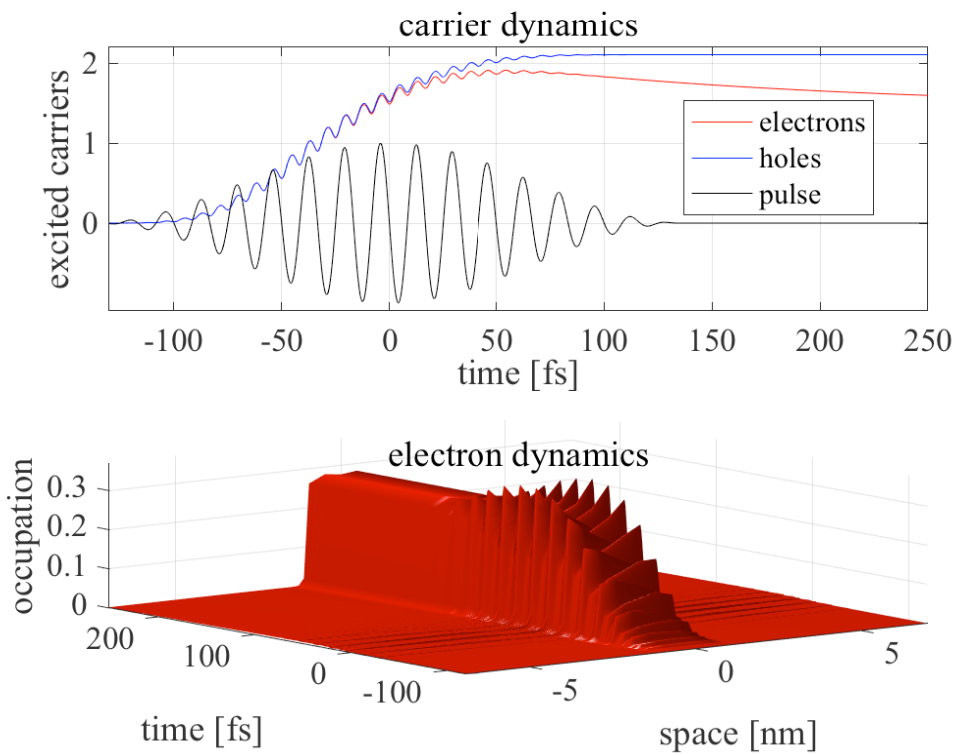

Fig. 2: Carrier dynamics in an optically excited nanowire of 2.2 nm length (5 sites). Electrons are emitted to free space resulting in an overall loss of carriers. additionally, the optical field of the exciting pulse is displayed for comparison. (pulse duration: 100fs, carrier wavelength: 5μm, field strength: 2.5 V/nm in free space, polarization: linear along the wire.)

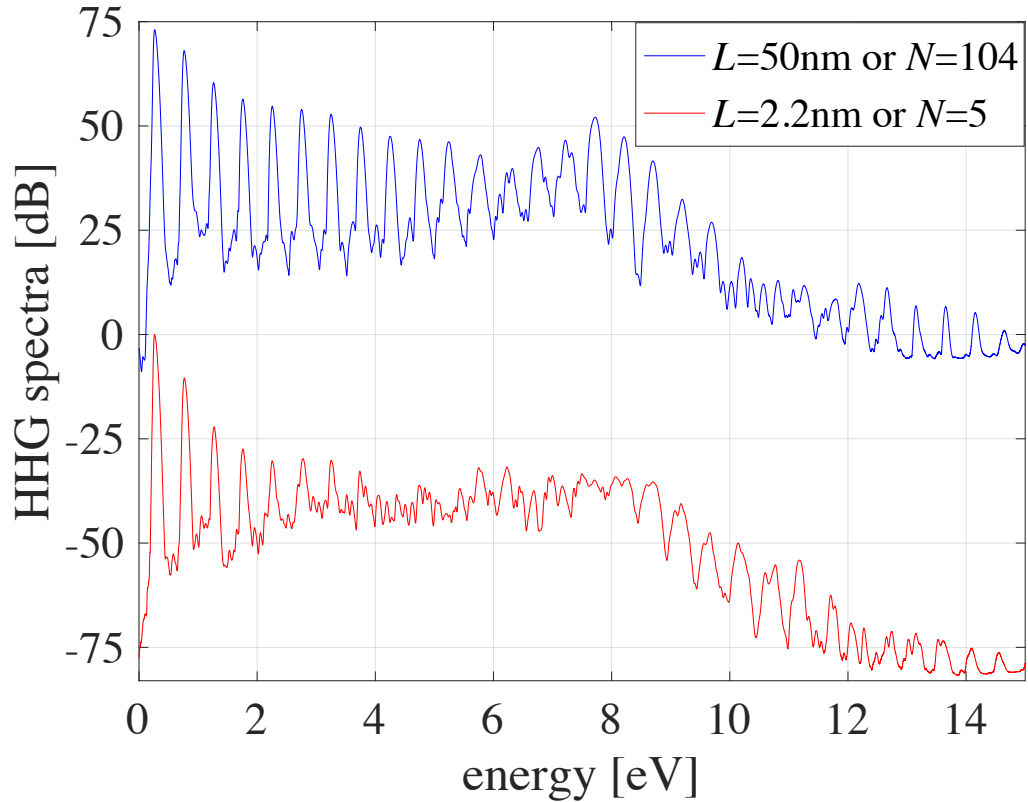

Fig. 3: Influence of confinement on HHG spectra. The spectra are generated by an intense pulse in a large bulk-like nanowire (blue line) and a small one (red line), for which quantum confinement plays an essential role (parameters as in Fig.2. The spectra are scaled by the squared number of sites. The blue line is shifted by +75dB for better visibility)

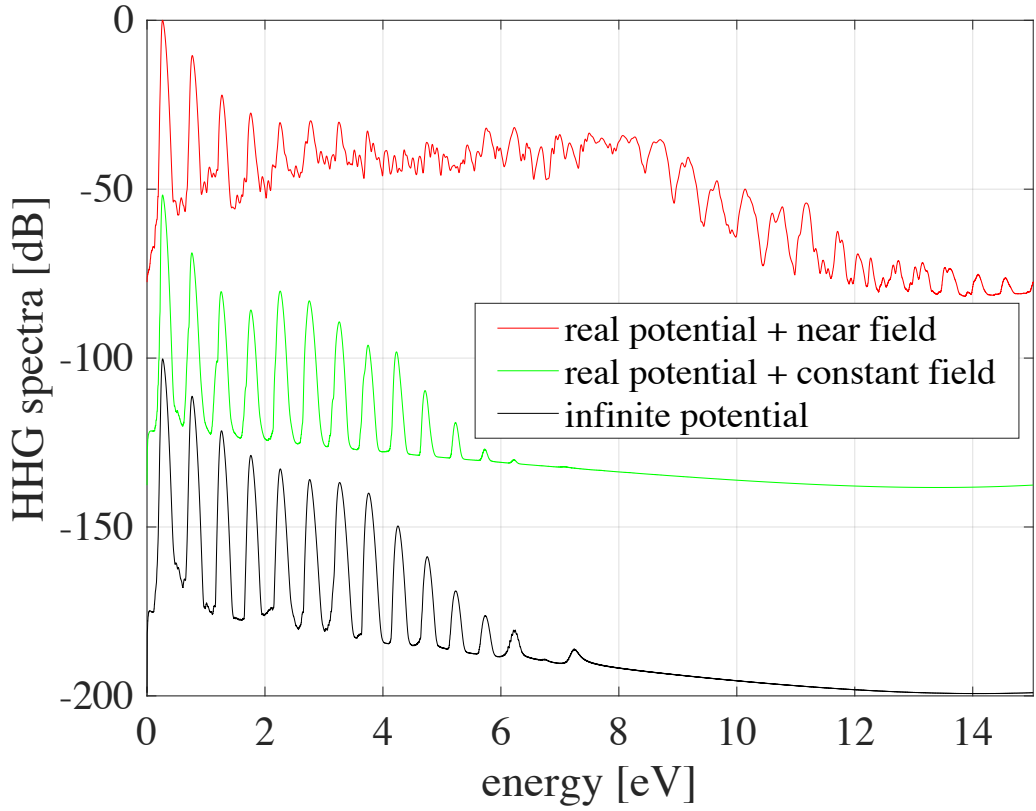

Fig. 4: Influence of the embedding environment on the HHG spectrum of a tiny nanostructures (5 sites or 2.2nm). Spectra of nanostructures with different confinement potentials and surrounding field structure are displayed (parameters as in Fig.2), The green and the black line are shifted by -50 and -100dB, respectively.

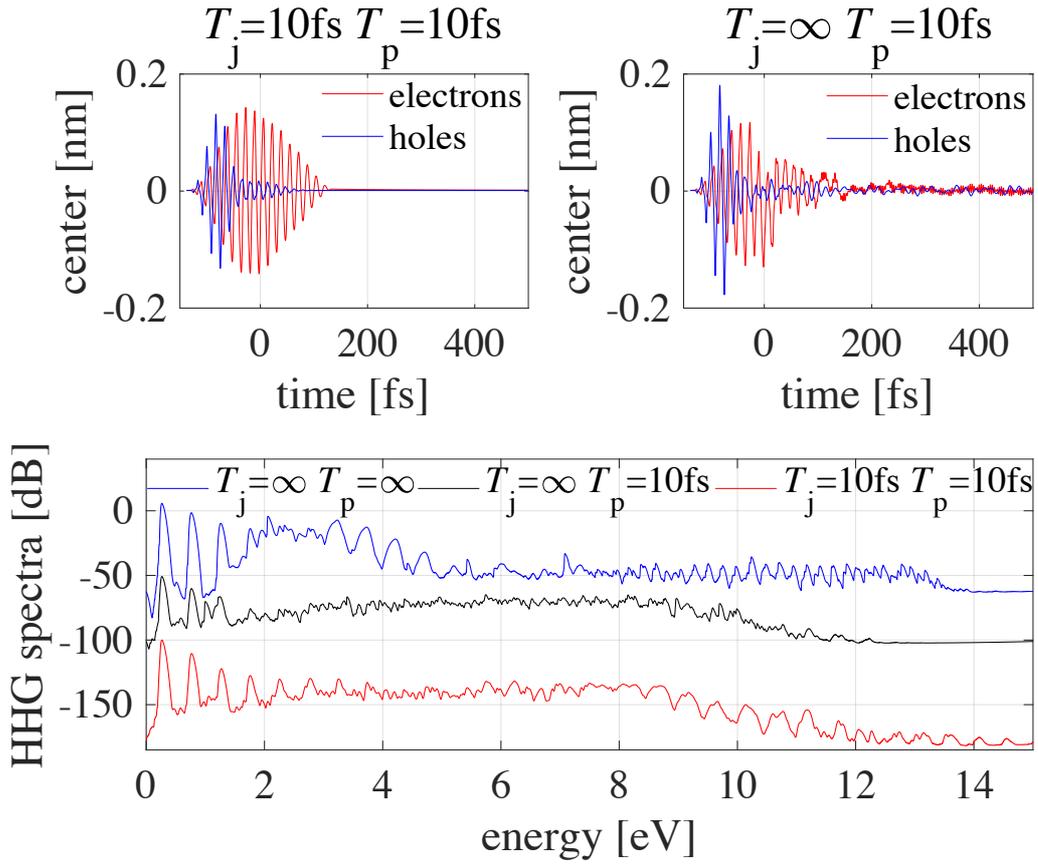

Fig. 5: Influence of damping (5 sites or 2.2nm) on the carrier evolution (upper row) and HHG in a nanostructure (lower graph). The black and the red line in the lower graph are shifted by -50 and -100dB, respectively.